\documentclass[twocolumn,pra,aps]{revtex4}\usepackage{psfrag}
\usepackage{epsfig}
\usepackage{amsmath}
\usepackage{amssymb}
\usepackage{bbm}
\usepackage{pifont}
\usepackage{pstricks,pst-node}
\usepackage{psfrag}
\newcommand{\bra}[1]{\ensuremath{\langle#1|}}
\newcommand{\ket}[1]{\ensuremath{|#1\rangle}}

\newcommand{\be}{\begin{equation}}
\newcommand{\ee}{\end{equation}}

\newcommand{\avg}[1]{\ensuremath{\langle #1 \rangle}}

\newcommand{\im}{\text{i}}
\newcommand{\adop}{\hat a^{\dagger}}
\newcommand{\aop}{\hat a}
\newcommand{\ain}{\hat a_{\text{in}}}
\newcommand{\adin}{\hat a^\dagger_{\text{in}}}
\newcommand{\aout}{\hat a_{\text{out}}}

\newcommand{\bdop}{\hat b^{\dagger}}
\newcommand{\bop}{\hat b}
\newcommand{\bin}{\hat b_{\text{in}}}

\newcommand{\hc}{\text{H.c.}}

\newcommand{\ketcat}{\ensuremath{|\Psi \rangle}}

\newcommand{\ie}{{\it i.e. }}
\newcommand{\eg}{{\it e.g. }}

\begin{document}

\title{Toward Quantum Superposition of Living Organisms}

\author{Oriol Romero-Isart$^1$}
\author{Mathieu L. Juan$^2$}
\author{Romain Quidant$^{2,3}$}
\author{J. Ignacio Cirac$^1$}
\affiliation{$^1$Max-Planck-Institut f\"ur Quantenoptik,
Hans-Kopfermann-Strasse 1,
D-85748, Garching, Germany.}
\affiliation{$^2$ICFO-Institut de Ciencies Fotoniques, Mediterranean Technology Park,
Castelldefels, Barcelona, 08860 Spain}
\affiliation{$^3$Instituci\'o Catalana  de Recerca i
Estudis Avan\c cats, E-08010, Barcelona, Spain}

\begin{abstract}
The most striking feature of quantum mechanics is the existence of superposition states, where an
object appears to be in different situations at the same time.
The existence of such states has been tested with small objects,
like atoms, ions, electrons and photons~\cite{Zoller05}, and even with molecules~\cite{Arndt99}.
More recently, it has been possible to create superpositions of collections of photons~\cite{Deleglise08}, atoms~\cite{Hammerer08}, or Cooper pairs~\cite{Friedman00}. Current progress in optomechanical systems may soon
allow us to create superpositions of even larger objects, like micro-sized mirrors or cantilevers~\cite{Bouwmeester03,Kippenberg08, Marquardt09, Karrai09}, and thus to test quantum mechanical phenomena at larger scales.
Here we propose a method to cool down and create quantum superpositions of the motion of sub-wavelength, arbitrarily shaped dielectric objects trapped inside a high--finesse cavity at a very low pressure. Our method is ideally suited for the smallest living organisms, such as viruses, which survive under low vacuum pressures~\cite{Rothschild01},
and optically behave as dielectric objects~\cite{Dziedzic87}. This opens up the possibility of testing the quantum nature of living organisms by creating quantum superposition states in very much the same spirit as the original Schr\"odinger's cat ``gedanken" paradigm~\cite{Schrodinger35}. We anticipate our essay to be a starting point to experimentally address fundamental questions, such as the role of life and consciousness in quantum mechanics.
\end{abstract}

\maketitle

The ultimate goal of quantum optomechanics is to push the motion of macroscopic objects towards the quantum limit, and it is a subject of interest of both fundamental and applied science~\cite{Kippenberg08, Marquardt09, Karrai09}. The typical experimental setup consists of an optical cavity whose resonance frequency  depends on the displacement of some mechanical oscillator. The mechanical motion shifts the resonance frequency,  and, consequently, the radiation pressure exerted into the mechanical object.
The overall effect yields the optomechanical coupling which should enable us to cool down to the ground state the mechanical motion~\cite{WilsonRae07, Marquardt07,Genes08}. We are currently witnessing an experimental race to reach the ground state using different setups, such as, nano- or microcantilevers~\cite{Groeblacher09a}, membranes~\cite{Thompson08}, or vibrating microtoroids~\cite{Schliesser08}.  It is expected that the achievement of the ground state will open up the possibility to perform fundamental and applied experiments involving quantum phenomena with these macroscopic objects, as pioneered by the works~\cite{Mancini97, Bose97, Armour02, Bouwmeester03}.

\begin{figure}
\centering
\includegraphics[scale=0.9]{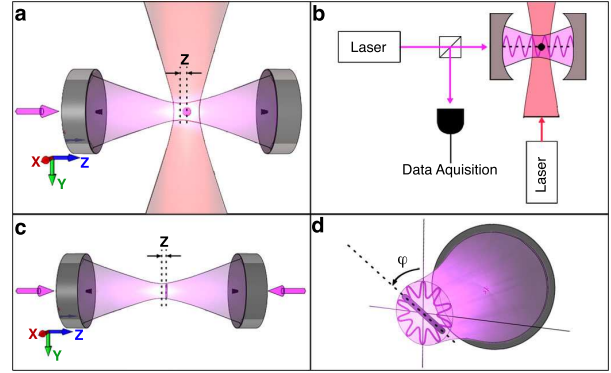}
\renewcommand{\figurename}{Fig.}
\caption{\textbf{Quantum optomechanics with dielectric objects trapped inside a high-finesse optical cavity}. a) A dielectric sphere is trapped by optical tweezers inside a high-finesse optical cavity. The confinement of the center-of-mass motion along the $z$-axis is harmonic with frequency $\omega_t$. The driving field generates a radiation pressure able to cool down the mechanical motion to the ground state. b) Experimental setup for trapping and cooling of dielectric spheres using two lasers, one for the driving and one for the trapping. c) The center-of-mass motion of a dielectric rod can also be trapped and cooled. In this case we assume self-trapping achieved by using two laser modes, see \ref{sec:trapping}. d) The rotational motion of a dielectric rod can also be cooled by generating a standing wave in the azimuthal angle. This can be achieved by superimposing two counterrotating Laguerre-Gauss modes.}\label{figure1}
\end{figure}

In this article, we propose dielectric objects levitating inside the cavity as new quantum optomechanical systems.
The fact that those are not attached to other mechanical objects avoids the main source
of heating, which is present in other optomechanical systems, and thus, should facilitate the
achievement of ground state cooling. Once this is achieved, we propose to create quantum
superpositions of the center of mass motional state of the object by sending a light pulse to the cavity simultaneously pumped with a strong field.
One of the main features of this proposal is that it 
applies to a wide variety of new objects and, in particular, to certain living organisms. Therefore, our proposal paves the path for the experimental test of the superposition principle with living creatures.

We consider an object with mass $M$, volume $V$, and  relative dielectric constant $\epsilon_r \neq 1$, which 
may be non--homogeneous. The object is trapped inside a cavity, either by an external 
trap, provided, for instance, by optical tweezers~\cite{Ashkin86} (Fig.~1a), or 
by self-trapping using two cavity modes (see \ref{sec:trapping} for details). The trap
is harmonic, so that the center of mass effectively decouples from any relative degree of freedom.
Along the cavity axis, this requires the size of the object to be smaller than the optical wavelength which 
is used for trapping and cooling.  The center-of-mass displacement, $z$, is then 
quantized as $\hat z= z_m(\bdop +\bop)$,  where  $\bdop$($\bop$) are creation(annihilation) phonon operators, 
and $z_m=(\hbar/2M\omega_t)^{1/2}$ is the ground state size, with $\omega_t$ the trap frequency. 
The resonance frequency of the optical cavity $\omega^0_c$ is modified by the presence of 
the dielectric object inside the cavity. A crucial relation is the frequency dependence on 
the position of the dielectric object, which can be estimated using perturbation 
theory (see~\ref{sec:frequency}). This position dependence gives 
rise to the typical quantum optomechanical coupling,
\be \label{eq:OMHamiltonian}
\hat H_{\rm OM} = \hbar g (\bdop+\bop) (\adop+ \aop)\, .
\ee
Here, $\adop$($\aop$) are the operators that create(annihilate) a resonant photon in the cavity. 
The quantum optomechanical coupling $g$ can be written as $g=\sqrt{n_{\rm ph}} g_0$, where 
$n_{\rm ph}$ is the number of photons inside the cavity and $g_0=z_m \xi_0$ ($\xi_0$ comes from 
the resonance frequency dependence on the position, see \ref{sec:coupling}). The enhancement of $g_0$ by a factor of $\sqrt{n_{\rm ph}}$ has been experimentally used 
to achieve the strong coupling regime in recent experiments with cantilevers~\cite{Groeblacher09b,Marquardt07,Dobrindt08}. 
Finally, the total Hamiltonian also includes the mechanical and radiation energy term as well as the 
driving of the cavity. See \ref{sec:Hamiltonian} for the details of these terms as well as 
the derivation of~\eqref{eq:OMHamiltonian}.

Besides the coherent dynamics given by the total Hamiltonian, there exists also a dissipative part 
provided by the losses of photons inside the cavity, parametrized by the decaying rate $\kappa$, and 
the heating to the motion of the dielectric object. Remarkably, our objects are trapped without 
linking the object to other mechanical pieces, and hence thermal transfer does not contribute to the mechanical damping $\gamma$. 
This fact constitutes a distinctive feature of our proposal, possibly yielding extremely high mechanical 
quality  factors. 
We have
investigated in detail the most important sources of
decoherence (see Appendix). First, heating due to coupling
with other modes, which have very high frequencies, is negligible when having a quadratic potential. Second, the maximum pressure required 
for ground state cooling is $\sim 10^{-6}\; {\rm Torr}$, which actually corresponds to the typical 
one used in  optomechanical experiments~\cite{Thompson08}.  The mechanical quality factor 
of our objects under this pressure is $\sim 10^9$, and it can be even increased in a higher vacuum. 
Third, blackbody radiation does not yield to a loss of coherence due to ``which-path" information at room and even much higher temperatures~\cite{Arndt99, Hackermuller04}.  Fourth, light scattering decreases the finesse of the cavity and produces heating. This sets the upper bound for the size of the objects in the current setup to be smaller than the optical wavelength. Fifth, the bulk temperature of the object remains close to the room temperature for sufficiently transparent objects at the optical wavelength, a fact that prevents its damage.

The rotational cooling of cylindrical objects, such as rods (see Fig.~1c), can also be 
considered. In this case, two counter-rotating Laguerre-Gauss modes can be employed 
to create a standing wave in the azimuthal angle $\phi$, as illustrated in Fig.~1d. The 
optomechanical coupling is then given by $g_0=(\hbar/2I\omega_t)\xi_0$, where $I$ is the moment of 
inertia. Using two modes, one can self-trap both the rotational and the center-of-mass translational 
motion, and cool either degree of freedom by slightly varying the configuration of the two modes 
(see the \ref{sec:coupling} for further details). Both degrees of motion can be simultaneously cooled 
if the trapping is provided externally (see Ref.~\cite{Bhattacharya07} for a proposal to cool the rotational motion of a mirror and a recent optomechanical experiment which uses a non-levitating nanorod \cite{Favero09}). 

Regarding the feasibility of our scheme, we require  the good cavity regime $\omega_t > \kappa$, in order 
to accomplish ground state cooling~\cite{WilsonRae07, Marquardt07,Genes08}.
Moreover, the strong coupling regime $g \gtrsim \kappa, \gamma$, is also required for quantum states 
generation.  Both regimes can be attained with realistic experimental parameters using dielectric spheres 
and rods. In particular, if one considers fused silica spheres of radius $250$ nm in a cavity with Finesse $10^5$ and length $4$ mm, one can get $g\sim \kappa \approx 2 \pi \times 180$ kHz, and $\omega_t \approx 2 \pi \times 350 $ kHz. See \ref{sec:numbers} for further details.

\begin{figure}
\centering
\includegraphics[scale=0.9]{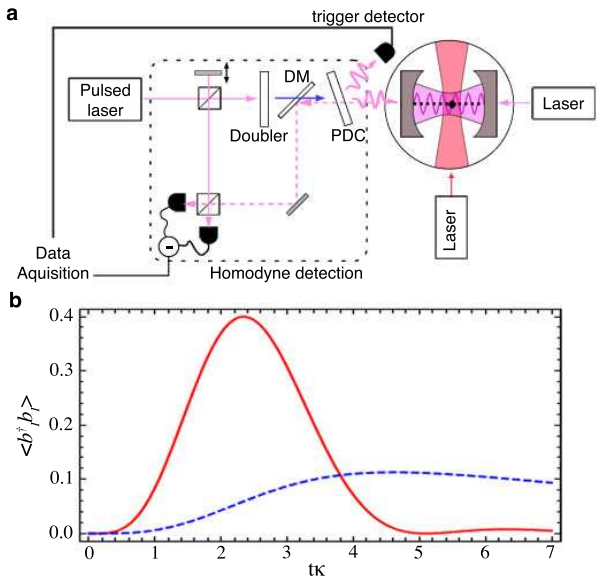}
\renewcommand{\figurename}{Fig.}
\caption{\textbf{Protocol to prepare quantum superposition states}. a) Experimental setup for implementing the protocol to prepare the quantum superposition state~\eqref{eq:superposition}. In the figure PDC stands for Parametric Down Conversion, and DM for Dichroic Mirror. A blue photon is converted into two red photons in the PDC.  One is detected and the other impinges to the cavity.
If it is reflected, the one photon pulse on top of the driving field goes back through the PDC (which is transparent),
then reflected downwards towards the homodyne detector by the DM. b) Mean number of phonons $\avg{\bdop_I \bop_I}$ imprinted to the mechanical oscillator by sending a one photon pulse to the cavity, see \ref{sec:superposition} for details. A gaussian pulse of width $\sigma = 5.6 \, \kappa$ is used. The red solid line corresponds to the strong coupling regime $g=\kappa$, whereas the blue dashed one corresponds to the weak coupling $g/\kappa=1/4$. In the strong coupling regime, the balanced homodyne measurement should be performed around the time where the mean number of phonons is maximum. This results in the preparation of the quantum superposition state \eqref{eq:superposition}. }\label{figure2}
\end{figure}

We tackle now the intriguing possibility to observe quantum phenomena with macroscopic objects. 
Notably, the optomechanical coupling \eqref{eq:OMHamiltonian} is of the same nature as the typical 
light-matter interface Hamiltonian in atomic ensembles~\cite{Hammerer08}. Hence, the same techniques 
can be applied to generate entanglement between gaussian states of different dielectric objects.

A more challenging step is the preparation of non-gaussian states, such as the paradigmatic quantum 
superposition state
\be \label{eq:superposition}
\ketcat=\frac{1}{\sqrt{2}}\left( \ket{0}+ \ket{1}\right)\, .
\ee
Here $\ket{0}$ ($\ket{1}$) is the ground state (first excited state) of the quantum harmonic 
oscillator. 
In the following, we sketch a 
protocol to create the state~\eqref{eq:superposition} ---see \ref{sec:superposition} for further details. The pivotal idea is to impinge the cavity with a single-photon
state, as a result of parametric down conversion followed by a detection of 
a single photon~\cite{Lvovsky01}. When impinging into the cavity, part
of the field will be reflected and part transmitted~\cite{Duan04}. In the presence of the red--detuned
laser, the coupling~\eqref{eq:OMHamiltonian} swaps the state of light inside the cavity 
to the mechanical motional state, yielding the entangled state  
$\ket{E}_{ab} \sim \ket{\tilde 0}_a \ket{1}_b + \ket{\tilde 1}_a \ket{0}_b $. Here $a$($b$) stands for the 
reflected cavity field(mechanical motion) system, and $ \ket{\tilde 0(\tilde 1)}_a$
is a displaced vacuum(one photon) light state in the output mode of the cavity. The protocol ends by performing a 
balanced homodyne measurement and by switching off the driving field. The motional state collapses 
into the superposition state $\ketcat=c_0 \ket{0} + c_1 \ket{1}$, where the coefficients 
$c_{0(1)}$ depend on the measurement result. See Fig.~2 for the experimental setup 
and results derived in \ref{sec:superposition}. This state can be detected by either 
transferring it back to a new driving field, and then performing tomography on the output field, 
or by monitoring the quantum mechanical oscillation caused by the harmonic trap. Moreover, the amplitude of the oscillation can be amplified by driving a blue-detuned field tuned to the upper motional sideband (see \ref{sec:superposition}).

A possible extension of the protocol is to impinge the cavity with other non-gaussian states, such as the NOON state or the Schr\"odinger's cat state $\ket{\alpha}+\ket{-\alpha}$~\cite{Ourjoumtsev07}, where $\ket{\alpha}$ is a coherent state with phase $\alpha$, in order to create other quantum superposition states. Furthermore, one can change the laser intensity dynamically to obtain a perfect transmission and avoid the balanced-homodyne measuremeny; any quantum state of light could be directly mapped to the mechanical system by the time-dependent interaction. Alternatively, one can tune the laser intensity to the upper motional sideband, so that a two-mode squeezing interaction is obtained in the cavity. In the bad-cavity limit (relaxing the strong coupling condition) one can use the entanglement between the output mode and the mechanical system to  teleport non-gaussian states~(O.R.I. {\em et al.}, in preparation).

\begin{figure}
\centering
\includegraphics[scale=.9]{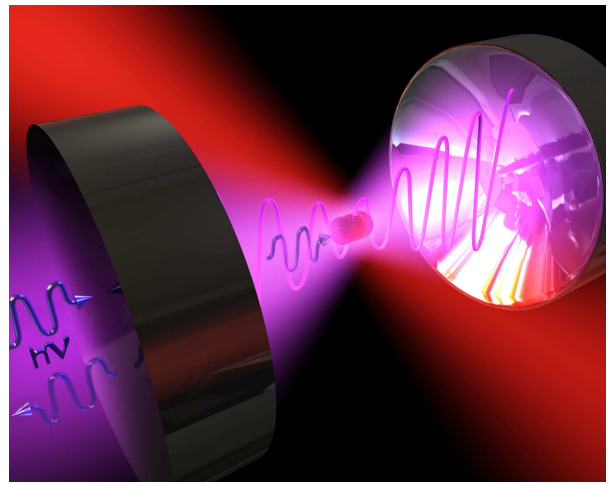}
\renewcommand{\figurename}{Fig.}
\caption{\textbf{Quantum superposition of living organisms}. Illustration of the protocol to create quantum superposition states applied to living organisms, such as viruses, trapped in a high-finesse optical cavity by optical tweezers.}\label{figure3}
\end{figure}

In the following we analyze the possibility of performing the proposed experiment with living organisms. 
The viability of this perspective is supported by the following reasons: i) living microorganisms  behave 
as dielectric objects, as shown in optical manipulation experiments in liquids~\cite{Dziedzic87}; 
ii) some microogranisms exhibit very high resistance to extreme conditions, and, in particular, to the 
vacuum required in quantum optomechanical experiments~\cite{Rothschild01}; iii) the size of some of 
the smallest living organisms, such as spores and viruses, is comparable to the laser wavelength, as 
required in the theoretical framework presented in this work; and iv) some of them present a transparency window (which prevents the damage caused by laser's heating), and still have sufficiently high refractive index.
As an example, the 
common influenza viruses, with size of $\sim100$ nm, can be stored for several weeks in vacuum 
down to $10^{-4}$ Torr~\cite{Greiff69}. In higher vacuum, up to $10^{-6}$ Torr, a good viability 
can be foreseen for optomechanics experiments.  Due to their structure (\eg lipid bilayer, nucleocapsid protein, and DNA), viruses present a transparency window at the optical wavelength which yields relatively low bulk temperatures~\cite{Steckl07}. Note that self-trapping or, alternative trapping methods, such as magnetic traps, could be used in order to employ lower laser powers.
The Tobacco Mosaic Virus (TMV) also presents 
a very good resistance to high vacuum~\cite{Rothschild01}, and has a rod-like appearance of $50$ 
nm wide and almost $1$ $\mu$m long. Therefore, it constitutes the perfect living candidate for 
the rotational cooling, see Fig.~1d. 

In conclusion, we have presented results that open the possibility to observe genuine quantum effects, 
such as the creation of quantum superposition states, with nano-dielectric 
objects and, in particular, with living organisms such as viruses, see Fig.~3. 
This entails the possibility to test quantum mechanics, not only with macroscopic objects, but also 
with living organisms. A direction to be explored is the extension to objects larger than the 
wavelength~(O.R.I. {\em et al.}, in preparation). This would permit to bring larger and more complex living organisms 
to the quantum realm; for instance,  the Tardigrade, which have a size ranging 
from $100$ $\mu$m $\sim$ $1.5$ mm~\cite{Nelson02}, and is known to survive during several days in 
open space~\cite{Jonsson08}. We expect the proposed experiments to be a first step to experimentally 
address fundamental questions, such as the role of life and consciousness in quantum mechanics, and maybe even implications in our interpretations of quantum mechanics~\cite{Simon09}.

We thank M. D. Lukin for discussions. We acknowledge funding by the Alexander von Humboldt foundation (O.R.I.),  European project SCALA,
the DFG --FOR635 and the excellence cluster Munich Advanced Photonics--,  Spanish Ministry of Sciences through Grants TEC2007-60186/MIC and CSD2007-046-NanoLight.es, Fundaci{\'o} CELLEX Barcelona, and Caixa Manresa.

\appendix{}
\section{Resonance frequency dependence on mechanical position} \label{sec:frequency}
Here we show how to estimate the frequency dependence on the mechanical coordinates of arbitrarily shaped dielectric objects.  Note that the resonance frequency $  \omega^0_{c} $, and the optical mode $\varphi_0 (\vec r)$ of the cavity without the dielectric object, are known solutions of the Helmholtz equation. The presence of the dielectric object, which is small compared to the cavity length, can be considered as a tiny perturbation on the whole dielectric present inside the cavity, and, thus, a perturbation theory can be used to estimate the resonance frequency
\be \label{eq:frequency}
\omega_{c}(q) \approx  \omega^0_{c} \left( 1 - \frac{\int_{V(q)} (\epsilon_r-1) \left | \varphi_0 (\vec r) \right |^2 d \vec r}{2  \int \left | \varphi_0(\vec r) \right |^2 d \vec r }   \right)\, .
\ee
Here $\epsilon_r$ is the relative dielectric constant of the object, and $V(q)$ is its volume at coordinates $q$. The integral in the numerator, which is performed through the volume of the object placed at coordinates $q$, yields the frequency dependence on $q$.

\section{Total Hamiltonian in quantum optomechanics} \label{sec:Hamiltonian}
The total Hamiltonian in quantum optomechanics can be typically written as
\be \label{eq:TotalHamiltonian}
\hat H_{\rm t}= \hat H_{\rm m}  + \hat H_{\rm OC} + \hat H_{\rm drive}\,.
\ee
The term $H_{\rm m}$ corresponds to the mechanical energy of the degree of motion $\hat q=q_{{\rm m}}(\bdop + \bop)$, which is assumed to be harmonically trapped. Therefore,  $\hat H_{\rm m}= \hbar \omega_t \bdop \bop$, where $\omega_t$ is the trapping frequency.  The driving of the cavity field, with a laser at frequency $\omega_L$  and strength $E$, related to the laser power $P$ by $|E|=\sqrt{2 P \kappa /\hbar \omega_{L}}$, is given by,
\be
\hat H_{\rm drive}=\im \hbar  \left ( E e^{-\im \omega_L t} \adop - E^\star e^{\im \omega_L t} \aop  \right) \,.
\ee

The last term corresponds to the radiation energy of the field inside the cavity $\hat H_{{\rm OC}}=\hbar \omega_c(\hat q) \adop \aop$, where $\adop$($\aop$) are the creation(annihilation) cavity photon operators.  When the equilibrium position, in the presence of the classical radiation pressure is at $q=0$, is fixed at the maximum slope of the standing wave inside the cavity, a linear dependence $\omega_c(\hat q)=\omega_c + \xi_0 \hat q$ is obtained, where $\omega_c=\omega^0_c+\delta$. The shift $\delta$ is caused by the equilibrium position of the dielectric object. See \ref{sec:coupling} for the specific quantities considering spheres and rods. Finally, it is convenient to perform a shift to the operators $\aop=\alpha + \aop'$ and $\bop=\beta +\bop'$ (the prime will be omitted hereafter), where $|\alpha|=\sqrt{n_{\rm ph}}$ is the square root of the number of cavity photons, and $\beta \approx - q_m \xi_0 |\alpha|^2 /\omega_t$. This transformation leaves invariant the dissipative part of the master equation (see \cite{WilsonRae07} for further details), and transforms the total Hamiltonian into $\hat H'_{\rm t}= \hat H_{\rm m} + \hat H_{\rm r}  + \hat H_{\rm OM}$, where $\hat H_{\rm r}= \hbar \omega_c \adop \aop$, and
\be
\hat H_{\rm OM} = \hbar |\alpha| q_m \xi_0  (\bdop+\bop) (\adop +\aop) \,.
\ee
 Note that one obtains that the optomechanical coupling is $g=|\alpha| g_0$, with $g_0=q_m \xi_0$. Note that the large term $|\alpha|$, which is typically of the order of $10^{4}$, compensates the small ground state size $q_m$.
\section{Protocol to create quantum superposition states} \label{sec:superposition}
Let us derive here the protocol to create quantum superposition phononic states of the type \eqref{eq:superposition}. We use the quantum Langevin equations and the input-output formalism. After going to the rotating frame with the laser frequency $\omega_L$, which is detuned to the resonance frequency by $\Delta=\omega_c-\omega_L$, displacing the photonic and phononic operators  $\aop = \alpha+ \aop'$,  $\bop = \beta+ \bop'$  (we will omit the  prime hereafter), choosing $\alpha \approx E /(\im \Delta + \kappa)$ and $\beta \approx g_0 |\alpha|^2/\omega_t$, so that the constant terms cancel, and neglecting subdominant terms, one obtains the quantum Langevin equation for the total Hamiltonian $\hat H'_{\rm t}$
\begin{eqnarray}
\dot a &=& - \left( \im \Delta  + \kappa \right) \aop  - \im g (\bdop + \bop)  + \sqrt{2 \kappa} \ain(t)  e^{\im \omega_L t}\, , \\
\dot b &=&   - \left( \im \omega_t  + \gamma \right)  \bop    -\im  g ( \adop  + \aop )+ \sqrt{2 \gamma} \bin(t)\, .
\end{eqnarray}

Note that one has the enhanced optomechanical coupling $g=|\alpha |g_0$. In the interaction picture, \ie $\aop_I = \aop e^{\im \Delta t}$ and $\bop_I= \bop e^{\im \omega_t t}$, if one chooses $\Delta= \omega_t$ (red-sideband), and perform the rotating-wave-approximation (valid for $\omega_t \gg g$),  derives the final equations
\begin{eqnarray}
\dot a_I &=& - \kappa \aop_I  - \im g  \bop_I   + \sqrt{2 \kappa} \ain(t)  e^{\im (\omega_L + \Delta) t} \\
\dot b_I &=&   -  \gamma  \bop_I    -\im g   \aop_I + \sqrt{2 \gamma} \bin(t) e^{\im \omega_t t }.
\end{eqnarray}

Next, we consider that the input for the photonic state is a light pulse with gaussian shape centered at the resonance frequency $\omega_c$, that is,
\be
\ket{\psi} = \int d \omega \phi(\omega) \ain^\dagger(\omega,L) \ket{\Omega}, \ee
where $\adin(\omega,L)$($\ain(\omega,L)$) are creation(annihilation) photonic operators out of the cavity at a distance $L$ and with frequency $\omega$, and $\phi(\omega)\propto\exp[-(\omega-\omega_c)^2/\sigma^2]$. Then, by recalling that $\ain (t,L) = \ain(t+L)=- \int d \omega e^{-\im \omega t} \ain(\omega,L)/\sqrt{2 \pi}$ ($c=1$), one has that $ \bra{\psi} \ain(t) \ket{\psi}  =0$ and $ \bra{\psi} \adin(t) \ain (t') \ket{\psi}  = \tilde \phi^\star (t-L) \tilde \phi (t'-L)$ (where $\tilde \phi(t)$ is the Fourier transform of $\phi(\omega)$). Solving the differential equations and obtaining $\bdop_I(t)$, one can compute $\avg{\bop_I(t)}$ (which is trivially zero since $\avg{\ain(t)}=\avg{\bin(t)}=0$), and $ \avg{ \bdop_I(t) \bop_I(t)}$.  The quantity $ \avg{ \bdop_I (t)\bop_I(t)}$ is plotted in Fig.~2b, for $g/\kappa=1$ and $g/\kappa=1/4$, with $\sigma=5.6 \, \kappa$, $\gamma \sim 0$, and $L=5/\kappa$. One can choose the width of the pulse so that half of the one-photon pulse enters into the cavity. Therefore, at some particular time $t^\star$ the entangled state
\be
\ket{E}_{ab} \sim \frac{1}{\sqrt{2}} \left(\ket{\tilde 0}_a \ket{1}_b + \ket{\tilde 1}_a \ket{0}_b \right)\, ,
\ee
is prepared. Here, $\ket{\tilde 0}_a$ ($\ket{\tilde 1}_a$) is the displaced vacuum (displaced one photon) state of the light system corresponding to the output field. This state yields $\avg{\bop_I(t^\star)}=0$ and $\avg{\bdop_I(t^\star) \bop_I(t^\star)}=1/2$, as obtained in Fig.~2b. The protocol finishes by performing the balanced homodyne measurement of the quadrature $\hat X_L(t) =(A^\dagger(t) + A(t))$ at time $t^\star$. Here $A(t)= \int_0^t \varphi(x,t) \aout(x,t)$ is the output mode of the cavity we are interested in, where $\varphi(x,t)$ can be computed. If one obtains the value $x_L$, the superposition state
\be
\ketcat_{b}=\frac{1}{\sqrt{2}} \left( c_0 \ket{0}_b + c_1 \ket{1}_b \right),
\ee
is prepared, where $c_{0(1)}=\bra{x_L}\tilde1(\tilde0)\rangle_a $. At the same time of the measurement, the driving field is switched off. 

Note that the distinguishability of the two orthogonal displaced states $\ket{\tilde 0} \pm \ket{\tilde 1}$ is exactly the same as for the non-displaced ones $\ket{0} \pm \ket{1}$.  However, the displacement of the output mode is of the order of $|\alpha|$ in the regime $\kappa \sim g$. This value, which is $\sim 10^4$ with the parameters proposed here, poses a challenge to the current precision of balanced homodyne detectors. This experimental challenge can be overcome by using alternative protocols (O.R.I. {\em et al.} in preparation). For instance, one can use a perfect transmission protocol which consists in using a time modulation of the optomechanical coupling $g(t)$, which can be implemented by varying the driving intensity, to perfectly transmit a particular light state  inside the cavity. Then, the beam-splitter interaction, given by the red-detuned driving field, perfectly transmits the input light state sent on top of the driving field to the mechanical system. The key feature of this protocol is that the balanced homodyne measurement is not required.

\textbf{Detection by amplification of the oscillation}. Let us assume that the state $(\ket{0}+\ket{1})/\sqrt{2}$ has been prepared in the mechanical system. The mean value of the position, in an harmonic trap, will oscillate with a frequency $\omega_t$ and amplitude proportional to the ground state size $q_m$. A coherent state would also oscillate with the same frequency. In order to distinguish both states, one could measure the fluctuations of the position, which for the superposition state will oscillate on time, whereas it will remain constant for the coherent state. This signal could be detected more easily by amplifying it by driving the cavity with a blue-detuned laser. One can find that the mean value of the position $\avg{\hat q (t)}$ under the influence of the two-mode squeezing interaction is given by
\be
\avg{\hat q (t)} = q_m \mu(t) \cos(\omega_t t)
\ee
where $\mu(t) = e^{-\kappa t/2} (\cosh(\chi t) + \kappa \sinh(\chi t)/2 \chi)$, with $\chi=\sqrt{g^2+\kappa^2/4}$ being a function which increases exponentially with $t$ and therefore, amplifies the oscillation.

\section{Self-trapping using two modes} \label{sec:trapping}
The self-trapping consists in using two optical modes $\aop_{1(2)}$, and combine them so that they provide trapping as well as the optomechanical coupling (in \cite{Hammerer09}, this configuration is also discussed in the context of optomechanics with cold atoms). The initial Hamiltonian in this case would be
\be
\hat H=\frac{\hat p^2_q}{2m} + \hbar \omega_{c,1} (q) \adop_1 \aop_1 + \hbar \omega_{c,2} (q) \adop_2 \aop_2 + H_{\rm L}.
\ee
In the displaced frame, $\aop_{1(2)} = \alpha_{1(2)}+\aop'_{1(2)}$ (we omit the primer hereafter), where $|\alpha_{1(2)}|$ is the square root of the number of photons for the mode $1$(2). Then, by expanding the resonance frequency up to the second order around $q=0$, \ie $\omega_{c,1(2)}(q)= \omega_{c,1(2)} +  \omega_{c,1(2)}' q + \omega_{c,1(2)}'' q^2/2$, fixing the key condition $|\alpha_1|^2 \omega'_{c,1}= -|\alpha_2|^2 \omega'_{c,2}$, and neglecting subdominant terms, one obtains the Hamiltonian
\be
\hat H=\frac{\hat p^2_q}{2m} +  \frac{m \omega^2_t}{2} \hat q^2 + \hbar \sum_{i=1}^2 \left [ (\xi_i \adop_i \hat q  + \hc)+  \omega_{c,i} \adop_i \aop_i \right] +\hat H_{\rm L}.
\ee
We have defined $\omega_t= [\hbar (\omega''_{c,1} |\alpha_1|^2 + \omega''_{c,2} |\alpha_2|^2)/m]^{1/2}$ and $\xi_i = \omega'_i \alpha_i$. Thus, provided $\omega''_{c,1} |\alpha_1|^2 + \omega''_{c,2} |\alpha_2|^2 >0$, one has the desired self-trapping and optomechanical coupling with the help of the two modes.

\section{Optomechanical coupling and trapping} \label{sec:coupling}

We compute here the optomechanical coupling for the sphere, assuming external trapping, and for the rod using self-trapping. In the latter, we derive two configurations required for cooling either the center-of-mass translational motion or the rotational motion. We will use the resonance frequency dependence estimated in \eqref{eq:frequency}.

\subsection{Dielectric sphere}

Let us consider the case of having a dielectric sphere of volume $V$ and relative dielectric constant $\epsilon_r$, and a TEM $00$ mode in the cavity. Then, the dependence of the resonance frequency on the center-of-mass position $ \vec r=(x,y,z)$, which can be estimated using~\eqref{eq:frequency}, is given by
\be \label{eq:frequencysphere}
\frac{\omega_{c}(\vec r)}{\omega^0_c} \approx 1 - \frac{V(\epsilon_r-1)  \left[ W^2 - 2 (x^2+y^2)\right] \cos^2 (\omega^0_c z/c)}{\pi W^4 d}\, .
\ee
Here $W$ is the waist of laser at the center of the cavity, and $d$ the cavity length. We consider a confocal cavity, $W=\sqrt{\lambda d/2 \pi}$. The object is assumed to be placed close to the center of the cavity and that the radius of the sphere is smaller than the laser waist.

We suppose external trapping at $x_0=y_0=0$ and $z_0=c\pi/4 \omega_c^0$, with frequency $\omega_t$~\eqref{eq:trappingsphere}. Then, the optomechanical coupling is given by $g_0=\sqrt{\hbar/2\rho V \omega_t} \xi_0$, where $\xi_0=\partial_z \omega_c(\vec r)|_0$ can be computed using \eqref{eq:frequencysphere}, and reads
\be 
\xi_0=\frac{(\omega^0_c)^2 (\epsilon_r-1)V}{c  d \pi W^2}\,.
\ee
The shifted frequency is $\omega_c=\omega_c(\vec r_0) =\omega_c^0 + \delta$, with $\delta=-V\omega_c^0(\epsilon_r-1)/2d \pi W^2$.

The external trapping can be achieved by optical tweezers. For spheres of radius $R$, mass $M$, and relative dielectric constant $\epsilon_r$, one can obtain, in the Rayleigh regime, that the trapping frequency is given by~\cite{Ashkin86}
%
\be \label{eq:trappingsphere}
\omega_t^2=\frac{6}{\rho c}  \left(\frac{\epsilon_r - 1}{\epsilon_r + 2}\right) \frac{I_0}{W_0^2}\,,
\ee
where $W_0$ is the laser waist, and $I_0$ the field intensity.

\subsection{Dielectric rod}

\begin{figure}[t]
\centering
\psfrag{L}{$L$} \psfrag{R}{$R$} \psfrag{a}{$a$} 
\psfrag{x}{$x$} \psfrag{y}{$y$} \psfrag{z}{$z$} \psfrag{p}{$\phi$} 
\includegraphics[scale=0.5]{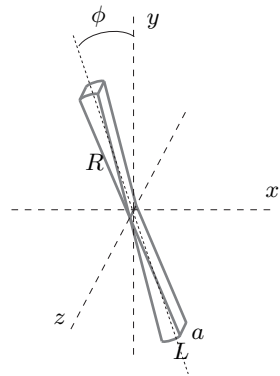}
\renewcommand{\figurename}{Fig.}
\caption{The model used as a rod in order to simplify the calculations. }\label{fig:rod}
\end{figure}

When considering a rod, in order to simplify the calculation, we model it as two opposed ``pieces of cake" of width $a$, arc $L$, and radius $R$, see Fig.~\ref{fig:rod}. Note that this corresponds to a small section of the waist of the laser, since we will take $R=W/2$. The volume of the rod is $V=RLa$, and its momentum of inertia $I=RLM/4\pi$, where $M$ is its mass. In the case of having a counterrotating Laguerre-Gauss (LG) mode $10$ and $-10$, the frequency dependence on the rotational angle $\phi$ and center-of-mass $z$ position (see Fig.~1c,d in the Letter) is given by
\be \label{eq:frequencyrodone}
\frac{\omega_{c,1}(\phi,z)}{\omega_{c}^0} = 1- \frac{ V (\epsilon_r-1) C_1 \cos^2 \left[ \omega^0_c z/c \right]   \cos^2 \left[ \phi \right] }{\pi  W^2 d },
\ee
with $C_1= 2 \left( 2 \sqrt{e} -3 \right)/\sqrt{e}$. In case of having a superposition of the LG modes $20$ and $-20$, one obtains the similar result
\be \label{eq:frequencyrodtwo}
\frac{\omega_{c,2}(\phi,z)}{\omega_{c}^0} = 1- \frac{ V (\epsilon_r-1) C_2 \cos^2 \left[\omega^0_c z/c \right]   \cos^2 \left[2 \phi \right] }{\pi  W^2 d },
\ee
with $C_2=\left( 8 \sqrt{e} -13 \right)/2 \sqrt{e} $. The rod is assumed to be placed close to the center of of the cavity and that its width $a$ is much smaller than the cavity length.

We propose the self-trapping configuration (see section \ref{sec:trapping}) in order to trap both center-of-mass translation and rotation, using, as before, the superposition of LG modes $10$ and $-10$ for the mode-$1$, and the superposition $20$ and $-20$ for the mode-$2$.

In case of aiming at cooling the translational motion, the equilibrium position is obtained at $\phi_0=0$ and $z_0=c\pi/8 \omega_c^0$, by translating the mode-$1$ a distance $z_0=c\pi/4 \omega_c^0$ with respect to the mode-$2$.  Then, using \eqref{eq:frequencyrodone} and \eqref{eq:frequencyrodtwo}, one can compute the $z$-optomechanical coupling  $g^z_0 = \sqrt{\hbar /2 \rho V \omega_{t,z}} \xi_0^z$, with $\xi_0^z=\partial_z \omega_{c,1}(\phi,z)|_0$, which reads
\be
\xi_0^z = -\frac{(\omega_c^0)^2 C_1(\epsilon_r-1)V}{c \sqrt{2} d \pi W^2}
\ee
One also has that $g^\phi_0=0$. The trapping frequency for the translation along the $z$-axis, $\omega_{t,z}$, and for the rotation $\omega_{t,\phi}$, can be computed using  $\partial^2_{zz} \omega_{c,1(2)}(\phi,z)| _0$, and $\partial^2_{\phi \phi} \omega_{c,1(2)}(\phi,z)| _0$. The shifted frequency is $\omega_{c,1(2)}=\omega_{c,1(2)}(\phi_0,z_0)=\omega_c^0 + \delta_{1(2)}$, with $\delta_{1(2)}=-V\omega_c^0(\epsilon_r-1) C_{1(2)} \cos^2(\pi/8)/d \pi W^2$.

In case of cooling the rotational motion, the equilibrium position is obtained at $\phi_0=7 \pi/12$ and $z_0=0$ by rotating the mode-$1$ an angle $\pi/4$ with respect to the mode-$2$.  Then, one can compute the $\phi$-optomechanical coupling $g^\phi_0=\sqrt{\hbar/2I\omega_{t,\phi}}\xi^\phi_0 = \partial_\phi \omega_{c,1}(\phi,z)|_0$, which reads
\be
\xi^\phi_0= -\frac{\omega_c^0 C_1 \sqrt{3}(\epsilon_r-1)V}{2 d \pi W^2}.
\ee
Also,  $g^z_0=0$. The trapping frequencies can be computed as in the translational motion coupling, but at the equilibrium position used for rotational cooling. The shifted frequency is in this case $\omega_{c,1(2)}=\omega_{c,1(2)}(\phi_0,z_0) =\omega_c^0 + \delta_{1(2)}$, with $\delta_{1(2)}=-3V\omega_c^0(\epsilon_r-1) C_{1(2)}/4d \pi W^2$.

Finally, let us mention that by a trapping provided externally, for instance, by means of optical tweezers, one could place the rod at the maximum slope of both the translational and azimuthal standing wave. Then, one would get $g^\phi_0 \neq 0$ and $g^z_0 \neq 0$ at the same time, and hence, one could cool both degrees of freedom simultaneously provided the trapping is tight enough.

\section{Heating and decoherence due to gas pressure} \label{sec:heating}

\subsection{Heating rate and mechanical damping}

Let us analyze here the heating and damping of the mechanical motion of the center-of-mass of a dielectric sphere due to the impact of air molecules inside the vacuum chamber. The air molecules of mass $m$ have mean velocity  $\bar v = \sqrt{3 K_b T/m}$, where $T$ is the temperature of the chamber, assumed to be at room temperature, and $K_b$ is the Boltzmann's constant.  The pressure inside the vacuum chamber is $P$, and the dielectric sphere has mass $M$, radius $R$, and is harmonically trapped with frequency $\omega_t$.

One can hence consider the Harmonic oscillator with additive white noise:
\be
\ddot z + 2 \gamma \dot z +  \omega_t^2 z = \xi(t) \,.
\ee
The stochastic force $\xi(t)$ describes the impact of air molecules. For white noise one has that $\avg{\xi(t)\xi(t')} =2 M^2 D \delta(t-t')= 4 K_b T M \gamma\delta(t-t')$ (thus  $D= 2 K_b T \gamma/M$), where in the last equation we have used the fluctuation dissipation theorem. The variance of the position can be computed solving the differential stochastic equation and supposing $\omega \gg \gamma$ (always fulfilled in our levitating spheres), one gets
\be
\avg{[z(t)-\avg{z(t)}]^2} \approx \frac{D}{2 \gamma \omega^2} \left \{ 1- e^{-2 \gamma t}\right\} \,.
\ee
By considering the equipartition principle, the variance allows us to compute the increase of energy by taking $\Delta E(t)= M \omega^2 \avg{[z(t)-\avg{z(t)}]^2} $. Hence one can compute the time $t^\star$ required to increment one quantum $\hbar \omega$ of energy in the quantum harmonic oscillator. This time should be larger than the inverse of the laser cooling rate $\Gamma$, which is defined as the time required to decrease one quantum of energy. The time $t^\star$ is given by solving $\Delta E(t^\star)=\hbar \omega$ and reads
\be
t^\star =  - \frac{1}{ 2 \gamma} \log \left( 1- \frac{ \hbar \omega}{ K_b T}\right) \approx \frac{\hbar \omega}{K_b T 2 \gamma}\,,
\ee
where we have used $\hbar \omega \ll K_b T$. Then the condition for ground state cooling is given by
\be \label{eq:GScondition}
t^\star \Gamma =   \frac{ \Gamma \hbar \omega}{K_b T 2 \gamma} \gg 1\,.
\ee
We determine now an expression for $\gamma$ which will depend on the properties of the gas surrounding the harmonic oscillator. We will derive it through kinetic theory. Assume our sphere is moving with velocity $v$. At the reference frame where the sphere has velocity equal to zero, one can compute the decrease of momentum of the sphere by the balance of momenta, given by the impact of one third of the particles colliding from behind with a velocity $\bar v - v$, minus those colliding in front with  velocity $v+\bar v$.
This can be written as
\begin{eqnarray}
\frac{\Delta p}{\Delta t} &=& (\bar v -v) \pi R^2 \frac{\rho}{3 m} 2 m (\bar v -v)-  (\bar v +v) \pi R^2 \frac{\rho}{3 m} 2 m (\bar v +v) \\ &=& - \frac{4 \pi R^2 \rho \bar v}{3 M} 2 M v = - \gamma 2 M v\,,
\end{eqnarray}
Using that the pressure of the gas is related to the density by $\rho= 3 P/\bar v^2 $, one obtains that
\be
\gamma= \frac{4 \pi R^2 P }{ M \bar v}\,.
\ee
Thus, inserting the value of gamma in~\eqref{eq:GScondition}, one finds an upper bound for the pressure required inside the vacuum chamber
\be
P \ll \frac{3 M \Gamma \hbar \omega}{8 m \bar v \pi R^2} \sim \Gamma \times 10^{-12}\; {\rm Torr} \; {\rm Hz}^{-1}.
\ee
We have used the spheres described in section \ref{sec:numbers}, and that the mass of molecules of air is $m \sim 28.6 $ u and $T=300 K$. Recalling that the typical cooling rate is of the order of hundreds of kHz~\cite{WilsonRae07, Marquardt07,Genes08}, one obtains the typical pressures of order $10^{-6}$ Torr used in experiments. With this pressure we have a damping of the order of mHz, which leads to extremely good mechanical quality factors of the order of $10^9$.

\subsection{Decoherence of the superposition state}

The same process of heating due to collisions of air molecules causes decoherence of a superposition state. Following~\cite{Joos85}, the relevant quantity is the localization rate
\be
\Lambda= \frac{3 m \bar v P}{\hbar^2} \pi R^2,
\ee
where we take the effective cross section as $\pi R^2$. This describes the decoherence  $\rho(x,x',t)=\rho(x,x',0)\exp[-\Lambda t(x-x')^2]$ due to scattering of air molecules. In the case of having the superposition state $\ket{0}+\ket{1}$ of the harmonic oscillator, the decoherence rate would be then given by $\Gamma_{{\rm dec}} = \Lambda z_m^2$, where $z_m=\sqrt{\hbar/2M\omega_t}$ is the ground state size. Recall that the heating rate is $\Gamma_+=1/t^\star=  K_b T 2 \gamma/\hbar \omega_t$, where $\gamma= 4 \pi R^2 P/ M \bar v$. Hence, using the expression of $\gamma$, one has that $\Gamma_{\rm dec}/\Gamma_+  \approx 1$, as it was to be expected for our harmonic oscillator.

\section{Bulk temperature}\label{sec:Bulktemperature}

In order to estimate the bulk temperature of the dielectric object attained after being heated by the lasers, we assume it behaves as a blackbody. Then, the steady state of the dielectric objects fulfills that the power absorbed
\be
P_{\rm abs}=\frac{\omega_L}{2} |E|^2 \Im[\alpha],
\ee
where $\omega_L$ is the laser frequency, $E$ the electric field, and $\alpha$ the polarizability,  
equals the power dissipated $P_{\rm rad}$ through black-body radiation
\be
P_{\rm rad}= A e \sigma \left[T^4-T^4_{\rm env} \right],
\ee
where $A$ is the area of the object, $e$ the emissivity ($\approx 1$), $\sigma$ the Stefan-Boltzmann constant, $T_{\rm env}$ the temperature of the vacuum chamber, and $T$ the bulk temperature. Thus, from $P_{\rm abs}=P_{\rm rad}$ one obtains the bulk temperature $T$. For a sphere of radius $R$ trapped by optical tweezers, this corresponds to
%
\be
T^4= I_0 \frac{4 \pi^3 R}{ e \sigma \lambda} \frac{3 \epsilon_2}{(\epsilon_1+2)^2+\epsilon_2^2} + T_{env}^4
\ee

where $P$ is the laser power and $\epsilon_r=\epsilon_1 + \im \epsilon_2$ the complex relative dielectric constant. Note that only here we have assumed $\epsilon_r$ to be complex, since $\epsilon_2$ is generally very small for the objects we consider.

\section{Experimental parameters for strong coupling and ground state cooling of dielectric spheres and rods} \label{sec:numbers}

We consider a confocal cavity of length $d=4$ mm, with a resonant laser at $\lambda=1064$ nm, which gives a waist at the center of the cavity of $W=\sqrt{\lambda d / 2 \pi} \approx 26.0 $ $\mu$m. If we assume a high-finesse optical cavity with $\mathcal{F}=10^5$, then the decaying rate is $\kappa=c\pi/2 \mathcal{F} d =2 \pi \times 188$ kHz. The presence of the sphere scatters photons out of the cavity and produces heating. A rough estimation, assuming that the total cross section is given by $\pi R^2$, sets an upper bound $\mathcal{F}= \pi W^2 /\pi R^2 \gtrsim 10^5$ for the radius of spheres of $\sim 80$ nm. A more rigorous calculation, using Mie theory, sets up the upper bound to $\sim 250$ nm (see Oriol Romero-Isart. {\em et al}. in preparation, where the effect of scattering of photons is studied in detail).

The dielectric objects are considered to be made of fused silica, with a density $\rho=2201$ Kg/${\rm m}^3$ and relative dielectric constant $\epsilon_r=2.1$. We take spheres of radius $250$ nm, and rods with length equal to the waist $W$, width $a=50$ nm, and arc length $L=50$ nm.

Using a laser of $1064$ nm, and a ratio $I_0/W_0^2=2 \, {\rm W}/\mu{\rm m}^4$, one has that the trapping frequency of the center-of-mass translation for the dielectric sphere provided by the optical tweezers is $\omega_t=2 \pi \times 351$~kHz (see \eqref{eq:trappingsphere}). Hence, $\kappa/\omega_t \sim 0.53$ places us well in the good cavity regime required for ground state cooling. On the other hand, the enhanced optomechanical coupling, with laser powers of $0.5$ mW, gives $g=2 \pi \times 182$ kHz, which also places us in the strong coupling regime $g \gtrsim \kappa, \gamma$.

Regarding the dielectric rod, for the translational motion cooling scheme, we achieve trapping frequencies of $\omega_{t,z}=2 \pi \times 552$ kHz, and $\omega_{t,\phi}=2 \pi \times 848$ kHz, and optomechanical coupling of $g_z=2 \pi \times 243$ kHz. For the rotational cooling, one has trapping frequencies $\omega_{t,z}=2 \pi \times 492$ kHz, $\omega_{t,\phi}=2 \pi \times 503$ kHz, and optomechanical coupling $g_\phi=2 \pi \times 276$ kHz. We assumed driving powers for the mode-$1$ of $4$ mW. In both cases, one gets the good cavity and strong coupling regimes.

Optical grade fused silica presents a very low absorption at $1064\; {\rm nm}$, with  $\epsilon_1=2.1$ and $\epsilon_2=2.5 \times 10^{-10}$. In these experimental conditions, the bulk temperature achieved for the dielectric spheres is estimated to be just around four degrees above the room temperature when using the optical tweezers.


\end{document}